\documentclass[preprint]{aastex}
\usepackage{color}
\usepackage{amsmath,graphics,graphicx}
\usepackage[suffix=]{epstopdf}
\usepackage{easybmat}
\usepackage{arydshln}
\usepackage{bm}
\usepackage{textcomp}
\usepackage{multirow}

\shorttitle{A new representation of the kappa-distribution}
\shortauthors{M. Battaglia et al.}

\newcommand{\lapprox} {\, \lower3pt\hbox{$\sim$}\lflap{\raise2pt\hbox{$<$}}\,}
\newcommand{\gapprox} {\, \lower3pt\hbox{$\sim$}\llap{\raise2pt\hbox{$>$}}\,}

\begin{document}
\title{Multi-thermal representation of the kappa-distribution of solar flare electrons and application to simultaneous X-ray and EUV observations}
\author{Marina Battaglia\altaffilmark{1,2}, Galina Motorina\altaffilmark{3,2}, and Eduard P. Kontar \altaffilmark{2}}
\altaffiltext{1}{Institute of 4D Technologies, School of Engineering, University of Applied Sciences and Arts Northwestern Switzerland, CH-5210 Windisch, Switzerland \\
Email: marina.battaglia@fhnw.ch}
\altaffiltext{2}{School of Physics \& Astronomy, University of Glasgow, G12 8QQ, Glasgow, Scotland, United Kingdom \\
Email:eduard.kontar@glasgow.ac.uk}
\altaffiltext{3}{Pulkovo Astronomical Observatory, Russian Academy of Sciences, Pulkovskoe sh. 65, St. Petersburg, 196140 Russian Federation 
Email: g.motorina@gao.spb.ru}

\begin{abstract}
Acceleration of particles and plasma heating is one of the fundamental problems in solar flare physics. An accurate determination of the spectrum of flare energized electrons over a broad energy range is crucial for our understanding of aspects such as the acceleration mechanism and the total flare energy. Recent years have seen a growing interest in the kappa-distribution as representation of the total spectrum of flare accelerated electrons. In this work we present the kappa-distribution as a differential emission measure. This allows for inferring the electron distribution from X-ray observations and EUV observations by simultaneously fitting the proposed function to RHESSI and SDO/AIA data. This yields the spatially integrated electron spectra of a coronal source between less than 0.1 keV up to several tens of keV. The method is applied to a single-loop GOES C4.1 flare. The results show that the total energy can only be determined accurately by combining RHESSI and AIA observations. Simultaneously fitting the proposed representation of the kappa-distribution reduces the electron number density in the analysed flare by a factor of $\sim 30$ and the total flare energy by a factor of $\sim$ 5 compared with the commonly used fitting of RHESSI spectra. The spatially integrated electron spectrum of the investigated flare between 0.043 keV and 24 keV is consistent with the combination of a low-temperature ($\sim$ 2 MK) component and a hot ($\sim$ 11 MK) kappa-like component with spectral index 4, reminiscent of  solar wind distributions.
\end{abstract}


\section{Introduction}\label{intro} 
One of the brightest manifestations of solar flares is emission in X-rays and radio wavelengths due to the presence of non-thermal particles and hot plasma.
Observations in these wavelengths provide information on the electron distribution and thus allow for inferring key solar flare parameters such as total number of accelerated electrons and total flare energy. The standard approach to infer electron distributions from X-ray spectra is by means of forward fitting an assumed shape of the electron distribution, though methods for inversion of the X-ray spectrum by means of regularisation do exist \citep[see][for a review]{2011SSRv..159..301K}. Typical X-ray spectra observed with RHESSI \citep{2002SoPh..210....3L} are consistent with a near-isothermal Maxwellian distribution at energies up to $\sim$ 20 keV and a non-thermal tail above these energies \citep[e.g.][as a review]{2011SSRv..159..107H}, although the relative importance of the two components can be different from flare to flare and between flare phases. The most simple model for the mean electron flux spectrum comprises of a Maxwellian:
\begin{displaymath}
F(E)\sim~E \exp(-E/k_BT) 
\end{displaymath}
plus a power-law:
\begin{displaymath}
F(E)\sim E^{-\delta}.
\end{displaymath}
 The major disadvantage of this model is that it requires the introduction of a low-energy cut-off of the non-thermal spectrum which is difficult to constrain observationally \citep[e.g.][]{Ho03,2005A&A...435..743S,2008SoPh..252..139K}. At the same time, due to the decreasing electron spectrum with energy (power-law index $\delta$ typically around 3 -- 6), the bulk of the electron energy is carried by supra-thermal electrons of energy around a few $k_BT$. Because of this, detailed knowledge of the electron spectra in the range of a few keV to some tens of keV is the key not only for understanding the physics of electron acceleration and transport but also for accurate estimates of total flare energies. 
Partially because of this, recent years have seen a growing interest in the kappa-distribution, i.e. a single analytic function that is close to a Maxwellian at low energies and has a power-law tail at higher energies:
\[
F(E)\propto E(1+E/k_BT_\kappa(\kappa-1.5))^{-(\kappa+1)}
\]
For energies much larger than the reference energy $k_BT_\kappa$, this is a power-law and for $\kappa \rightarrow \infty$ it approaches a Maxwellian distribution. Kappa-distributions are supported by theoretical considerations of particle acceleration in collisional plasmas. \citet{2014ApJ...796..142B} showed that distributions close to a kappa-distribution can be formed in a stochastic acceleration scenario 
- at least in the range of supra-thermal particles, while escape of electrons could modify the high energy tale of the distribution. Indeed, some X-ray flare spectra observed with RHESSI can be fitted with kappa-distributions \citep{2009A&A...497L..13K, 2013ApJ...764....6O, 2015ApJ...799..129O}, most readily events that consist of a single source at all energies. However, the fit developed by \citet{2009A&A...497L..13K} assumes thin-target emission and only considers electron-ion bremsstrahlung. At the same time, supra-thermal electrons, via non-thermal recombination,
emit free-bound emission \citep{1969MNRAS.144..375C,1970MNRAS.151..141C,2010A&A...515C...1B,2012A&A...539A.107D}, whose contribution to the total X-ray spectrum can be significant. Inclusion of free-bound emission leads to an increase of the total X-ray flux from the same number of electrons thus reducing the electron number required to produce the observed emission. 

The RHESSI X-ray spectrum provides high spectral resolution information on the possible shape of the electron distribution in the keV up to MeV energy range, but
RHESSI is only sensitive down to X-ray energies above $3$~keV (or even higher, depending on the attenuator state). Because of this, lower electron energies 
are weakly constrained. Combining X-ray observations with EUV observations from SDO/AIA \citep{2012SoPh..275...17L} allows for determining electron distributions over a much wider energy range from below $0.1$~keV
up to the deca-keV range. \cite{2013ApJ...779..107B} first showed combined observations of electron distributions inferred from fits to RHESSI X-ray spectra with electron distributions from differential emission measures (DEM) observed with SDO/AIA, treating the SDO/AIA and RHESSI  observations as independent datasets. An improvement over this would be simultaneous fitting of RHESSI and AIA data. \citet{2014ApJ...789..116I} performed such joint fitting on the low energy (thermal) component of microflares, but fitted any existing high-energy X-ray component separately. 

In this paper we: a) develop and test a fitting model that represents the kappa-distribution and includes a complete treatment of relevant emission mechanisms (free-free, free-bound, double photon, and line emission) based on CHIANTI 7.1 \citep{1997A&AS..125..149D,2013ApJ...763...86L}. Our model describes the electron distribution as a sum of Maxwellians, weighted by a coefficient proportional to the differential emission measure, so that the total emitting, volume averaged, distribution represents the kappa-distribution; b) present a method that allows for true simultaneous fitting of the whole RHESSI spectrum and AIA data with the new model by combining the AIA temperature response and the RHESSI temperature response into one response matrix.
We test fitting the new model to RHESSI spectra alone and simultaneously with AIA on a single-loop solar flare which was previously analysed by \citet{2013ApJ...779..107B}.
The results of fitting RHESSI flare spectra alone show that, for a typical flare, the number of electrons inferred using the new model
is significantly less than that from a standard kappa-distribution. This implies that a much smaller number of supra-thermal electrons than previously thought is present in coronal sources.
The combined analysis of RHESSI and SDO/AIA data confirms the result from the RHESSI fitting alone, but an additional DEM component is needed to account for the low-temperature emission observed by AIA to which RHESSI is not sensitive. This suggests that a single multi-thermal kappa-distribution fits a relatively broad range of electron energies apart from the range near $\sim 0.1$~keV which is indicative of the additional contribution of coronal $1-2$~MK 
plasma.

The paper is structured as follows. In Section 2 we describe the proposed DEM, show that it indeed represents the kappa-distribution, and explain how the electron distribution is inferred from it. In Section 3 we present results from fitting actual RHESSI spectra with the new model. Section 4 explains the method that is used for simultaneous fitting of RHESSI and AIA data and applies it to actual flare observations. The findings are discussed in Section 5. 
\section{Description of kappa-distributions via differential emission measure}
Distributions of electrons that have a thermal core and a power-law tail, in some cases can be described by a kappa-distribution \citep[e.g.][]{1968JGR....73.2839V, 2009A&A...497L..13K}:
\begin{equation} \label{eq:kasp}
f_\kappa(E)=n_\kappa\frac{2\sqrt{E}}{\sqrt{\pi(k_B T_\kappa)^3}}\frac{\Gamma(\kappa+1)}{\Gamma(\kappa-0.5)(\kappa-1.5)^{3/2}}\left(1+\frac{E}{(\kappa-1.5)k_B T_\kappa}\right)^{-(\kappa+1)}
\end{equation}
Here, $T_{\kappa}$ is the mean kinetic temperature of the electrons defined via their mean energy  $\langle E\rangle=(3/2) k_B T_{\kappa}$ and $k_B$ is the Boltzmann constant. Assuming an isotropic distribution, the variable change $f_\kappa(E)dE=4\pi v^2f_\kappa(v)dv$ results in an electron velocity flux distribution as used by \citet{2014ApJ...796..142B}:
\begin{equation}\label{eq:f_v3}
f_\kappa(v) =n_\kappa \left(\frac{1}{\pi \kappa \theta ^2}\right)^{3/2}
\frac{ \Gamma (\kappa +1) }{\Gamma (\kappa -1/2)}
\left(1+\frac{v^2}{\kappa \theta ^2}\right)^{-\kappa -1},
\end{equation}
where the characteristic speed $\theta ^2 = \frac{2k_BT_{\kappa}}{m_e}\frac{\kappa -1.5}{\kappa}$ is used and $n_\kappa=\int f_\kappa(v)d^3v$ is the number density associated with the accelerated electron distribution. 
The parameters of the kappa-distribution have a clear physical meaning.
In the stochastic acceleration scenario in a collisional plasma of \citet{2014ApJ...796..142B}, the characteristic speed is related to the thermal velocity of the Maxwellian electron distribution on which acceleration is acting, leading to the formation of a tail whose spectral index $\kappa$ is determined via balance between diffusive
acceleration and collisions. 
\subsection{Kappa-distribution and differential emission measure}
\begin{figure}
\vspace{-20pt}
  \begin{center}
  \includegraphics[width=10cm]{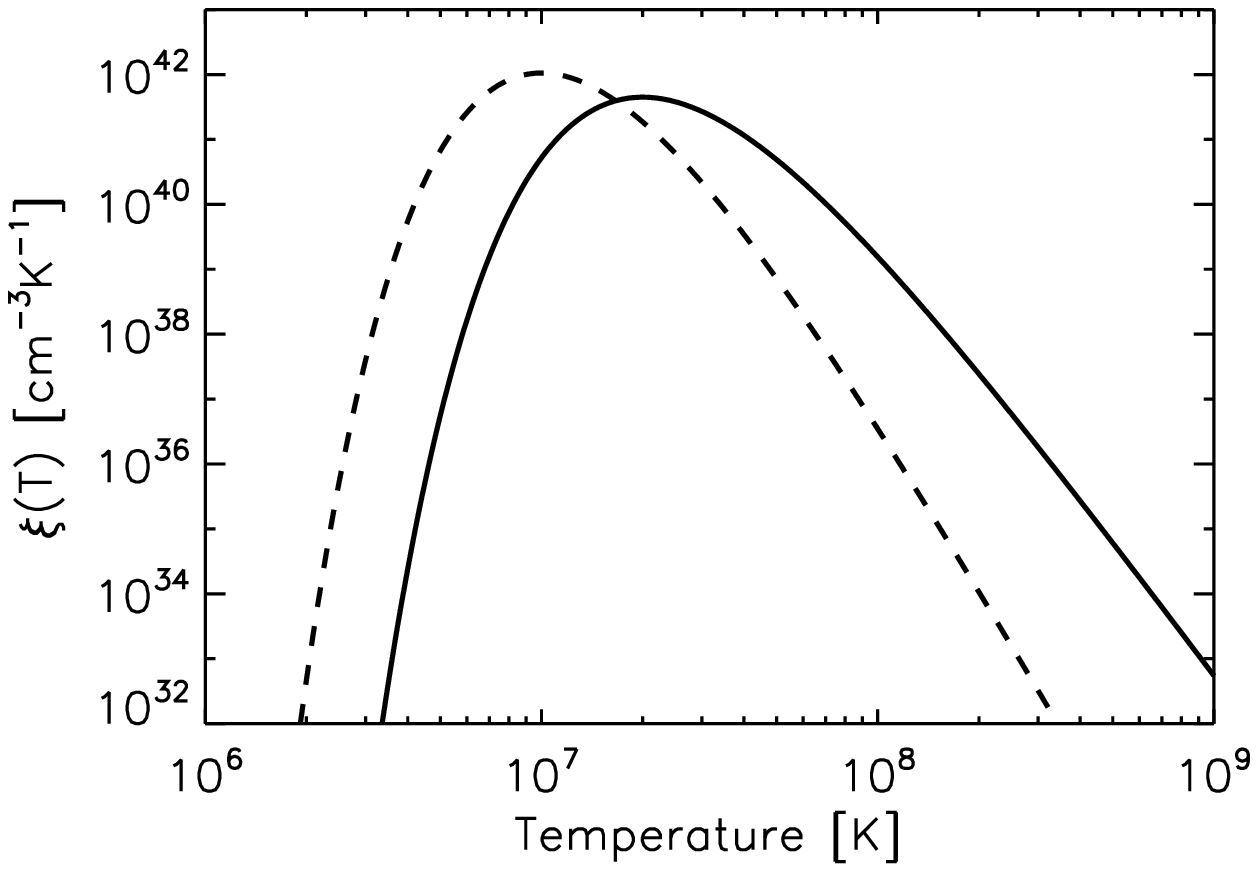}
  \end{center}
\vspace{-20pt}
\caption{Differential emission measure, $\xi (T)$, for plasma with emission measure $EM=10^{49}$~cm$^{-3}$ and two sets of parameters:
 $\kappa =6.5$, $T_{max}=20$~MK (solid line); $\kappa =8.5$, $T_{max}=10$~MK (dashed line).}
\vspace{-10pt}
  \label{fig:DEM}
\end{figure}

The mean electron flux spectrum $\langle nVF (E)\rangle=\int_Vn(r)F(E,r) dV$ in the emitting volume $V$ is the only observable that can be inferred directly from the X-ray spectrum without assumptions on the density or emitting volume. It can be related to the DEM $\xi(T)$ via the Maxwellian electron distribution at temperature $T$ \citep{1988ApJ...331..554B,2013ApJ...779..107B}:
\begin{equation} \label{eq:meaneflux}
\langle nVF(E)\rangle =\frac{2^{3/2}E}{(\pi m_e)^{1/2}}\int_0^{\infty}\frac{\xi(T)}{(k_BT)^{3/2}}\exp{(-E/k_BT)}\mathrm{d}T \quad \mathrm{[electrons~keV^{-1}s^{-1}cm^{-2}].}
\end{equation}
Therefore, knowing the differential emission measure, one can compute the mean electron flux spectrum in the emitting volume.
Here we introduce a DEM $\xi(T)$ of the shape:
\begin{equation}\label{eq:DEM_propto}
\xi (T) \propto T^{-(\kappa+0.5)}\exp \left(-\frac{T_\kappa}{T}(\kappa-1.5)\right) 
\end{equation}
This DEM drops with increasing $T$ since $\xi(T)\propto T^{-(\kappa+0.5)}$ for $T\gg T_\kappa$ and it falls off quickly for $T < T_\kappa$
due to $\exp (-T_\kappa/T)$. The DEM has a single maximum, $d\xi(T)/dT=0$ at $T_{max}= T_\kappa(\kappa-1.5)/(\kappa+0.5)$ (see Figure \ref{fig:DEM}).
It has the advantage that the integral over $\xi (T)$ can be solved analytically. Moreover, it represents the kappa-distribution given in Equations (\ref{eq:kasp}) and (\ref{eq:f_v3}). This can be shown as follows: 

The total emission measure $EM$ is defined as the integral of $\xi(T)$ over all temperatures in the plasma
\begin{equation}\label{eq:EM}
EM = \int_{0}^{\infty}\xi (T)dT \propto  \int_{0}^{\infty}T^{-(\kappa+0.5)}\exp \left(-\frac{T_\kappa}{T}(\kappa-1.5)\right) dT \quad \mathrm{[cm^{-3}]}.
\end{equation}
This integral can be solved using the gamma function $\Gamma (x)\equiv\int _{0}^{\infty}y^{x-1}\exp(-y)dy$ resulting in
\begin{equation}
EM=\frac{\Gamma (\kappa-0.5)}{(\kappa-1.5)^{(\kappa-0.5)}}T_\kappa^{-(\kappa-0.5)}
\end{equation}
Hence, one can write $\xi(T)$ as:
\begin{equation}\label{eq:DEM_T0}
\xi (T) =  \frac{EM(\kappa-1.5)^{(\kappa-0.5)}}{\Gamma (\kappa -0.5)T_\kappa} \left(\frac{T_\kappa}{T}\right) ^{\kappa+0.5}\exp \left(-\frac{T_\kappa}{T}(\kappa-1.5)\right) 
\end{equation}
i.e. the proportionality constant in Equation (\ref{eq:DEM_propto}) is $EM \cdot T_\kappa^{(\kappa -0.5)} (\kappa-1.5)^{(\kappa-0.5)}/\Gamma (\kappa -0.5)$. 

For a uniform plasma density $n=n_e=n_i$, the emission measure can be written as $EM=n^2V$, where $V$ is the emitting volume. Inserting Equation (\ref{eq:DEM_T0}) into Equation (\ref{eq:meaneflux}) one finds:
\begin{equation} \label{eq:meaneflux4}
\langle nVF(E)\rangle =n^2 V\frac{2^{3/2}}{(\pi m_e)^{1/2}(k_BT_\kappa)^{1/2}}
\frac{ \Gamma (\kappa +1) }{(\kappa-1.5)^{1.5}\Gamma (\kappa -1/2)}\frac{E/k_BT_\kappa}{(1+E/k_BT_\kappa(\kappa-1.5))^{\kappa +1}}.
\end{equation}
If we introduce an isotropic electron distribution function $\langle f(v)\rangle$, $n=\int \langle f(v)\rangle d^3v$, so that $F(E)dE=v\langle f(v)\rangle d^3v$,
and $d^3v=4\pi v^2 dv$ we can write:
\begin{equation} \label{eq:f_v}
\langle f(v)\rangle =\frac{m_e n}{4\pi v^2}\frac{2^{3/2}}{(\pi m_e k_BT_\kappa)^{1/2}}
\frac{\Gamma (\kappa +1) }{(\kappa-1.5)^{3/2}\Gamma (\kappa -1/2)}\frac{m_ev^2/(2k_BT_\kappa)}{[1+m_ev^2/(2k_BT_\kappa(\kappa-1.5))]^{\kappa +1}}
\end{equation}
where $E=m_ev^2/2$ was used. Or, simplifying:
\begin{equation}\label{eq:f_v2}
\langle f(v)\rangle =n \left(\frac{m_e }{2\pi k_BT_\kappa(\kappa-1.5)}\right)^{3/2}
\frac{ \Gamma (\kappa +1) }{\Gamma (\kappa -1/2)}\left(1+\frac{m_ev^2}{(\kappa-1.5)2k_BT_\kappa}\right)^{-\kappa -1}.
\end{equation}
This is identical to Equation (\ref{eq:f_v3}) when $T_\kappa$ is expressed via the characteristic speed. 

The DEM given in Equation (\ref{eq:DEM_T0}) thus indeed represents the kappa-distribution. It is defined by three parameters, EM, $T_{\kappa}$, and $\kappa$, which can readily be found by fitting X-ray and EUV spectra. It is implemented in OSPEX (called f\_multi\_therm\_pow\_exp.pro and henceforth referred to a $\xi_\kappa(T)$)\footnote{for documentation on OSPEX see http://hesperia.gsfc.nasa.gov/rhessi3/software/ \\ spectroscopy/spectral-analysis-software/index.html}.
Note that the function implemented in OSPEX does not give the parameters $\kappa$ and $T_\kappa$, directly, but $\alpha=\kappa+0.5$, and $T_{max}=T_\kappa(\kappa-1.5)/(\kappa+0.5)$. 
\section{Observations of a single-loop GOES C4.1 flare}
\begin{figure}
\centering
\includegraphics[width=12cm]{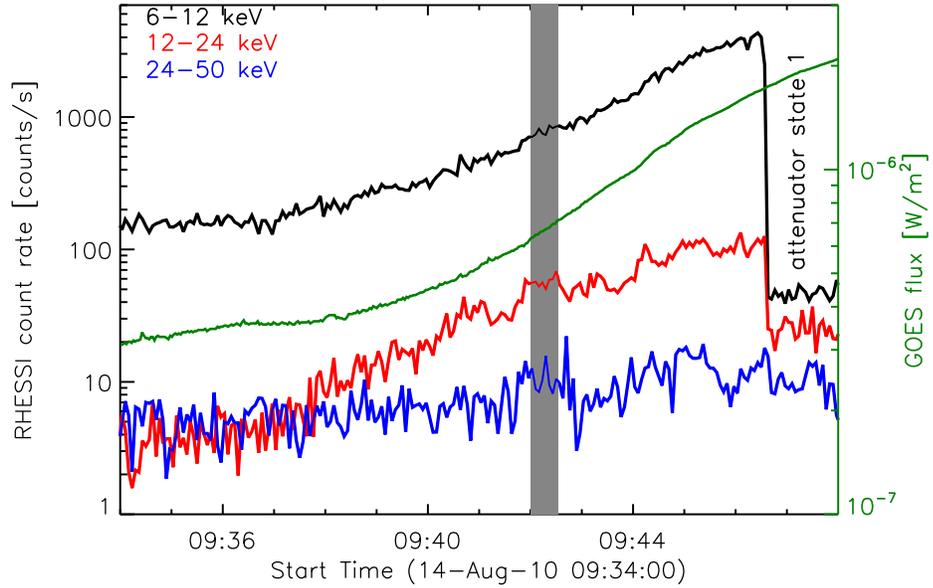}
\caption{RHESSI lightcurves of SOL2010-08-14T10:05 at 6-12 keV (black), 12-24 keV (red), and 24-50 keV (blue). The grey bar gives the fitted time-interval. The green line is the GOES lightcurve.}
\label{fig:lc}
\end{figure}

We demonstrate fitting of the two different functions ($\xi_\kappa(T)$ versus the original f\_thin\_kappa.pro routine, henceforth referred to as \textit{thin\_kappa}) to one of the flares that were presented in Battaglia \& Kontar 2013. The flare (SOL2010-08-14T10:05) happened on August 14 2010 with HXR peak around 09:46 UT. It had a single loop morphology seen in both RHESSI and SDO/AIA with only faint footpoint emission at energies above $\sim$ 18 keV.  Thus it is ideally suited for a study of the electron distribution in a single source without having to resort to imaging spectroscopy. RHESSI light-curves at different energy bands are shown in Figure \ref{fig:lc} along with the GOES lightcurve.
\begin{figure}
\centering
\includegraphics[width=0.8\linewidth]{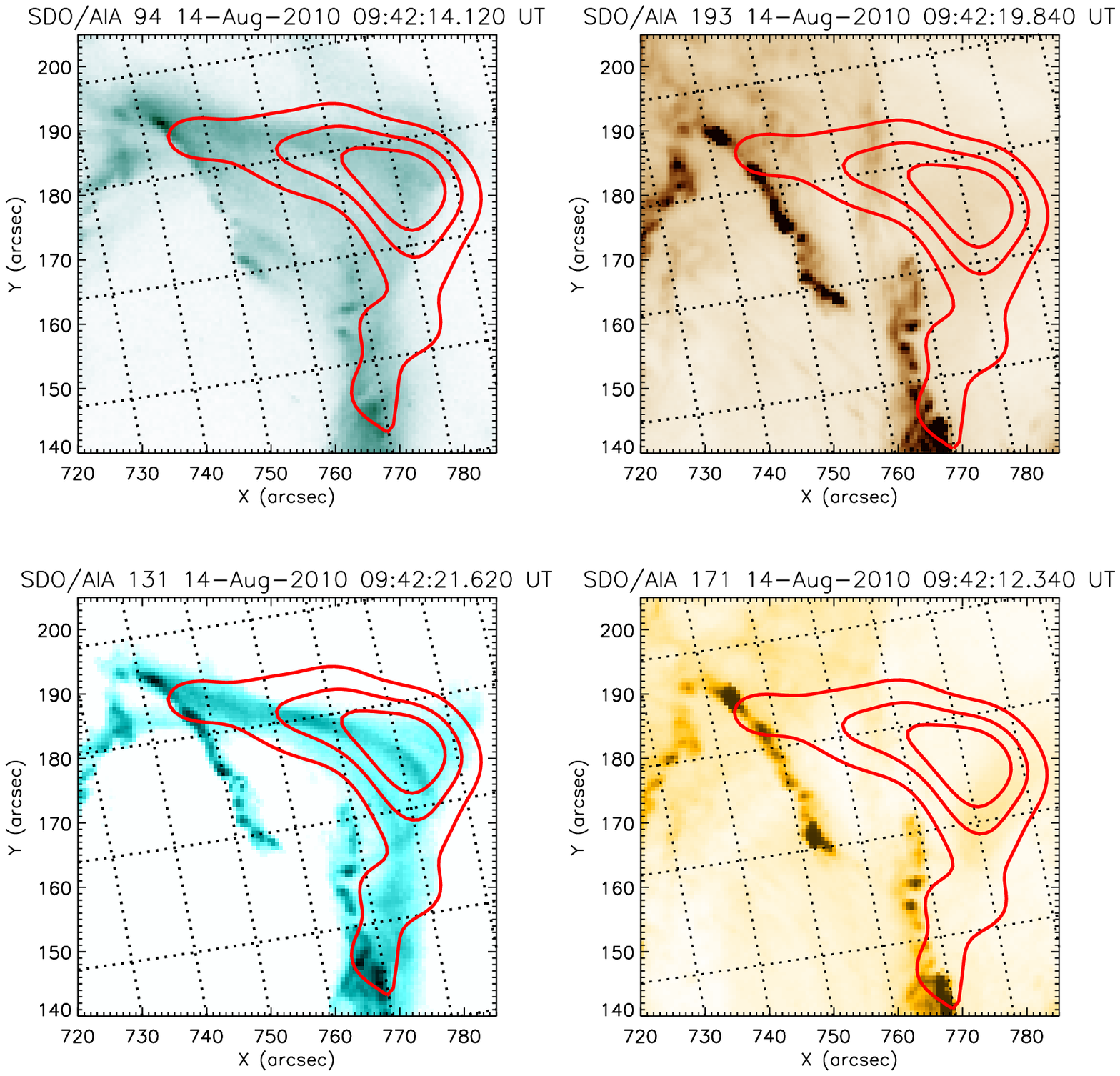}
\caption{AIA images in 4 wavelength-channels at the time for which the RHESSI spectrum was fitted. 30\%, 50\%, and 70\% contours from a RHESSI CLEAN image at 6-12 keV are given in red.}
\label{fig:aiaim}
\end{figure}
We analyze the time interval between 09:42:00 - 09:42:32 UT. This was in the rise phase of the flare where pile-up \citep{Smith02} was negligible and HXR emission was observed up to $\sim$ 24 keV. Figure~\ref{fig:aiaim} shows AIA images in four different wavelength bands overlaid with contours from RHESSI CLEAN \citep{Hur02}. The AIA images at 193 \AA\ and 171 \AA\ display two flare ribbons that are connected by a loop visible in 131 \AA\ and 94 \AA\ and are co-spatial with the RHESSI source at 6--12 keV. RHESSI spectra from detector 4 were fitted with $\xi_\kappa(T)$ and with \textit{thin\_kappa} for comparison. The fitted energy range was 7--24 keV. Figure \ref{fig:spectra} shows the spectra and fit results for the two different fit models.  The fit parameters are given in Table \ref{tab:param}. In the case of \textit{thin\_kappa}, an additional Gaussian line had to be fitted to account for the Fe-line complex near 6.7 keV. In the case of $\xi_\kappa(T)$, no Gaussian line is needed since line emission is included in the fit model. While the high-energy tail is very well constrained by the fit, a range of emission measures and temperatures can be fitted with similar $\chi^2$ for all models. For this reason, the fit was re-run several times for different starting parameters and using slightly different lower and upper limits of the fitted energy range ($\pm$ 1 keV) to get an estimate of the uncertainties. These uncertainties are given in Table \ref{tab:param} to indicate the range of likely fit parameters. \begin{figure}
\centering
\includegraphics[width=0.99\linewidth]{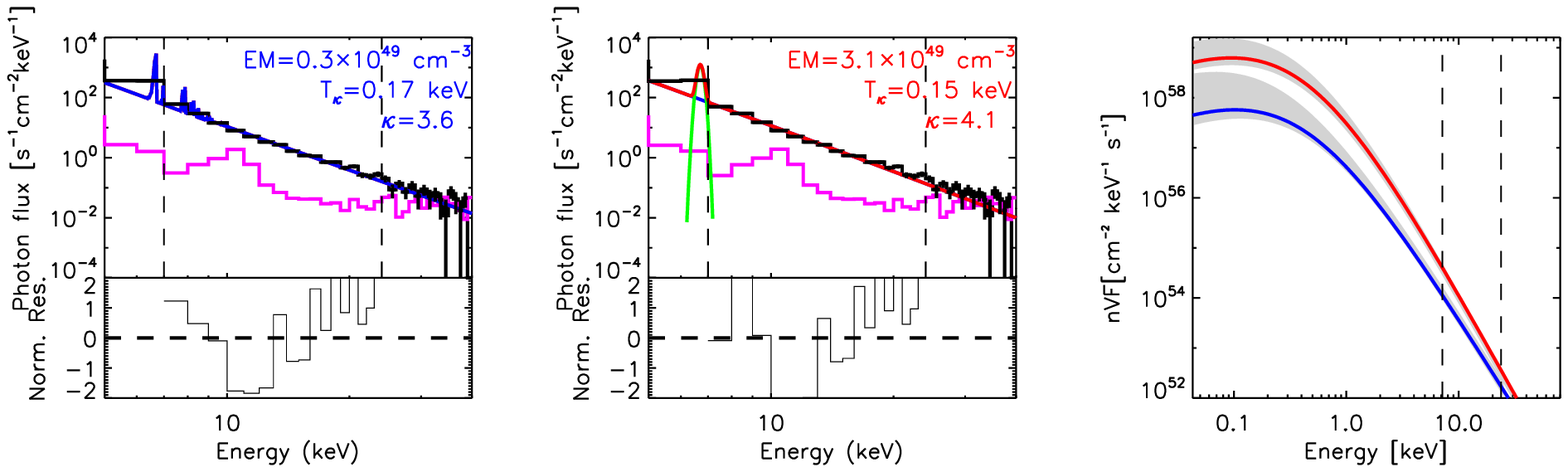}
\caption{Left and middle: RHESSI photon spectra (black points) with background level (purple histogram) and fitted model. Left: $\xi_\kappa(T)$ (blue line). Middle: \textit{thin\_kappa} (red line) plus Gaussian line (green). The dashed lines indicate the fitted energy range (7-24 keV). Right: Mean electron flux spectrum $\langle nVF(E) \rangle$ from \textit{thin\_kappa} (red) and $\xi_\kappa(T)$ (blue). The grey areas give the confidence range of the fit.}
\label{fig:spectra}
\end{figure}
For both models we calculated the mean electron flux distribution $\langle nVF(E) \rangle$. The comparison is shown in Figure \ref{fig:spectra}, right. It is apparent that the low-energy electron distributions are significantly different, a result of the ill-constrained emission measure and temperature. This illustrates the need for better constraints of the low-energy component such as can be obtained from AIA data.  
\section{Combined RHESSI and SDO/AIA data analysis}
RHESSI is sensitive to plasma temperatures higher than $\sim$ 8 MK, while combination of the six coronal AIA wavelength-channels (94 \AA, 131 \AA, 171 \AA, 211 \AA, 335 \AA, and 193 \AA ) covers the $\sim$ 0.5--16 MK range. Thus simultaneous analysis of RHESSI and AIA data improves the measurement of the low-temperature component of a given electron distribution and allows for extending the accessible energy range down to $\sim$ 0.1 keV. We developed a method with which RHESSI and AIA data can be fitted simultaneously by means of generating a single temperature response matrix that consists of the AIA temperature response and the RHESSI temperature response. We then apply the method using $\xi_\kappa(T)$ in the form of Equation (\ref{eq:DEM_T0}).
\subsection{Simultaneous fitting of RHESSI and AIA data}
\begin{figure}
\centering
\includegraphics[width=0.6\linewidth]{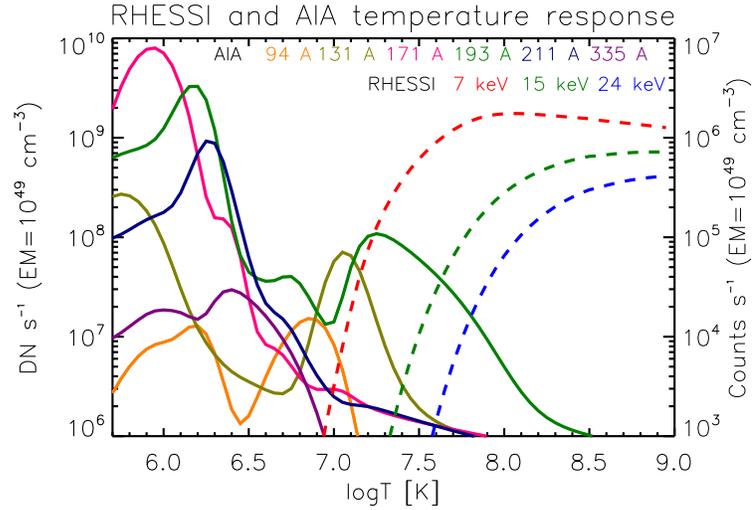}
\caption{AIA temperature response for 6 AIA wavelength-channels (solid lines, left axis) and RHESSI temperature response at 7 keV, 15 keV, and 24 keV (dashed lines, right axis), for an emission measure of $10^{49}$ cm$^{-3}$. }
\label{fig:newmatrix}
\end{figure}

The signal $g_i$ detected at a given  X-ray energy  or EUV wavelength is the DEM of the source multiplied by the detector response and the temperature contribution function:\begin{equation}\label{eq:g}
g_i=R_{ij} \xi_j dT_j
 \end{equation}
 where $\mathbf{g}=(\mathbf{g^{AIA}},\mathbf{g^{RHESSI}})$. $\mathbf{g^{AIA}}$ is a vector containing the EUV data numbers (DN/s) from the total flaring area in the 6 wavelength-channels. $\mathbf{g^{RHESSI}}$ is the observed RHESSI count rate spectrum (counts/s).  
 $\mathbf{R}$ is the combined temperature response matrix, including the SDO/AIA temperature response (for i=1,...6) and the RHESSI temperature response (for $i\ge 7$). The latter is constructed using the RHESSI spectral response \citep[SRM, see][]{Smith02} and the thermal bremsstrahlung radiation function for the plasma temperatures contained in the AIA temperature response. Figure \ref{fig:newmatrix} shows the AIA thermal response and the RHESSI thermal response for 7 keV, 15 keV, and 24 keV. The temperature coverage of $\mathbf{R}$ ranges from 0.043 keV (0.5 MK) to $\sim$ 86 keV (1000 MK).  The extension to 86 keV is necessary in order to fit the high-energy part of the spectrum properly. For this, the database files containing the tabulated thermal X-ray spectra had to be recalculated\footnote{These tables are calculated using the latest version of CHIANTI (7.1). They are part of solar soft and available through OSPEX. See also http://hesperia.gsfc.nasa.gov/rhessi3/software/spectroscopy/spectral-analysis-software/\\index.html}.  Since RHESSI is not sensitive to low temperatures, the RHESSI temperature response is set to zero for all energies that represent temperatures  $<{0.1}$ keV.  
 \subsection{Application of simultaneous fitting using $\xi_\kappa(T)$}
We now apply the method described above to the flare SOL2010-08-14T10:05 for the same time interval considered in Section 3 (09:42:00 - 09:42:32 UT) and test whether a single multi-thermal kappa distribution fits the extended energy range down to 0.043 keV. Since such fit capability is not built-in in OSPEX the standard IDL mpfit procedure was used for this part.
The uncertainties of the RHESSI data were calculated as 
$ C_{err}=\sqrt{(C+B)/L+B_{err}^2+(0.02 C)^2}$, where $C$  is the (background subtracted) count-rate, $B$  the background count-rate,  $B_{err}$ the statistical uncertainty of the background count-rate, and $L$ the detector live-time \citep{Smith02}. A systematic count-rate error of 2\% was assumed. The considered RHESSI energies were 7--24 keV, as in the analysis in Section {3}. For SDO/AIA, unsaturated images from six EUV wavelengths taken at around 09:42:15 UT were used.
AIA 131 \AA\ and 94 \AA\ wavelength channels show the flare loop co-spatial with the loop in RHESSI images (compare Figure \ref{fig:aiaim}). Thus we assume that both instruments observe the same emitting plasma and, for AIA, use the total DN within the region outlined by the RHESSI 50\% contour. The errors on the AIA data numbers were computed as $DN_{err}=\sqrt{DN+(0.2 DN)^2}$, including a systematic error of 20\% \citep[e.g.][]{2010ApJ...714..636L}. 
Fitting the whole temperature range with a single $\xi_\kappa(T)$ did not result in a satisfactory $\chi^2$, therefore a second $\xi_\kappa(T)$ was added and fitted simultaneously with the first. We call the two components $\xi_\kappa^{hot}(T)$ and $\xi_\kappa^{cold}(T)$ in the following. Figure 6 shows the results of the simultaneous fit of AIA and RHESSI data ($\chi^2=1.6$), separated into RHESSI and AIA data sets. It can be seen that $\xi_\kappa^{hot}(T)$ fits both data sets, while the cold component is only detectable with AIA but not with RHESSI. It dominates the emission in 171 \AA\ and contributes considerably to 211 \AA\ and to a lesser degree to 193 \AA. Also, for this cooler component, $\kappa$ is only weakly constrained and the other fit parameters (EM, T, and $\chi^2$) are not sensitive to its value. This suggests that this component is dominated by thermal foreground coronal emission at $\sim$ 1--2 MK, as has been suggested by \citet{2012ApJ...760..142B} and \citet{2014ApJ...780..107K} for different events.
The left panel in Figure \ref{fig:DEM_nVF_combined} shows a comparison of the DEMs found by fitting RHESSI spectra alone, from the simultaneous fit of RHESSI and AIA, and from AIA data alone, for which the DEM was inferred via the regularized inversion method developed by \citet{2012A&A...539A.146H}. In addition, the AIA loci-curves (i.e. observed data divided by the temperature response function), showing the temperatures for which AIA is most sensitive, are given. The AIA DEM drops sharply at the edges of the AIA temperature sensitivity range ($\log_{10}T=5.7$  and $\log_{10}T=7.5$). Further it has to be noted that the DEM from the combined fit does not have a "dip" around $\log_{10}T=6.6$ like the DEM found from regularized inversion of AIA data. This "dip" is a consequence of the DEM reconstruction since neither of the AIA wavelength channels that are used for the calculation of the DEM has a clear peak in the response around those temperatures \citep[][]{2013ApJ...779..107B}.

In summary, combined fitting of AIA and RHESSI data improves the reconstruction of the DEM overall. Instead of the continuously growing DEM from the fit to RHESSI data only, the DEM has a clear maximum around $\log_{10}T=6.8$ and a smaller peak around $\log_{10}T=6.1$. The resulting mean electron flux spectra $\langle nVF (E)\rangle$ are shown in the right panel of Figure \ref{fig:DEM_nVF_combined}. It can be seen that, by adding AIA data, i.e. by introducing additional constraints on low temperatures, the mean electron flux spectrum at low energies (up to $\sim$ 1 keV) is reduced by approximately one order of magnitude compared with $\langle nVF(E)\rangle$ from RHESSI data alone. This has far reaching consequences for the total energy of the flare which is dominated by the spectrum around 1 keV (see Discussion).
\begin{figure}
\centering
\includegraphics[width=0.99\linewidth]{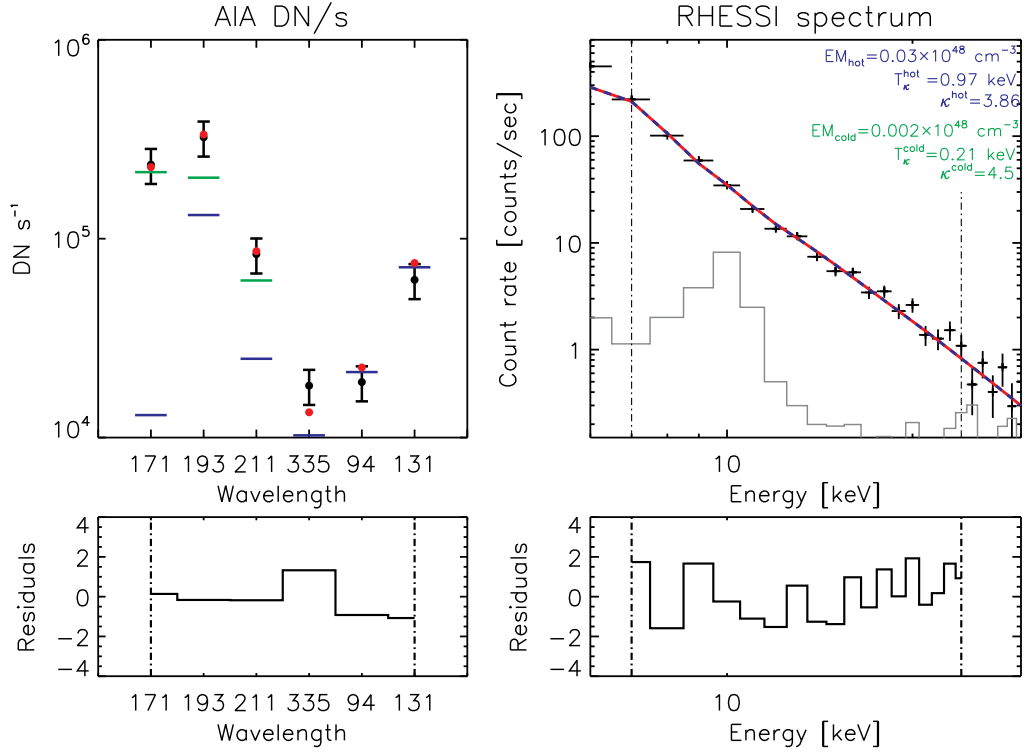}
\caption{Simultaneous fits of RHESSI and AIA data. Left: AIA DN as a function of wavelength-channel with the expected DN from the fit. The fit consisted of two $\xi_\kappa(T)$ ($\xi_\kappa^{hot}(T)$, blue lines and $\xi_\kappa^{cold}(T)$, green lines). The red dots gives the total. Note that $\xi_\kappa^{cold}(T)$ (green) only affects very low temperatures. It is therefore not visible in the RHESSI count spectrum but contributes significantly to the emission in AIA wavelengths 171 \AA, 193 \AA, and 211 \AA. Right: RHESSI count-rate spectrum.}
\label{fig:fittingRHESSI_SDO}
\end{figure}
\begin{figure}
\centering
\includegraphics[width=0.99\linewidth]{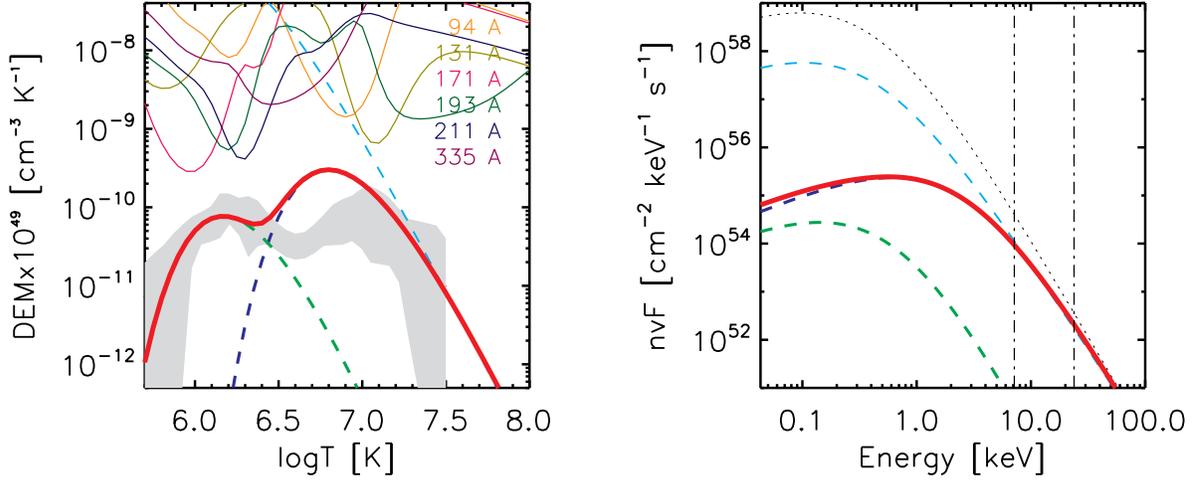}
\caption{Left: Comparison of DEMs from different methods: DEM from fit with one $\xi_\kappa(T)$ to RHESSI data (light-blue dashed); DEM from simultaneous fit of RHESSI and AIA with two $\xi_\kappa(T)$s ($\xi_\kappa^{hot}(T)$, blue dashed line and $\xi_\kappa^{cold}(T)$, green dashed line, compare Figure \ref{fig:fittingRHESSI_SDO}). The red line gives the sum of the two fits.  AIA loci-curves are indicated near the top of the plot. The grey area indicates the DEM (with confidence range) from AIA data, only, found by regularized inversion. Right: $\langle nVF(E)\rangle$ obtained from the simultaneous fit of AIA and RHESSI data (red). The dotted black line and dashed light-blue lines give $\langle nVF(E)\rangle$ from \textit{thin\_kappa} and from a single $\xi_\kappa(T)$ fitted to RHESSI data. 
}
\label{fig:DEM_nVF_combined}
\end{figure} 
\begin{deluxetable}{lcccccc} 
\tabletypesize{\footnotesize}
\tablecaption{Fit parameters from different fit models, electron number density $n$, flare energy density $U_\kappa$, and total flare energy $E_{tot}$.  Uncertainties are given as subscripts and superscripts. \label{tab:param}}
\tablecolumns{10}
\tablehead{ Fit model & 	EM $[10^{48}\mathrm{cm^{-3}}]$  & $\kappa$ & 	$T_\kappa$ [keV]&$n$ [$10^{10}$cm$^{-3}$] & $U_\kappa$ [erg cm$^{-3}$]& $E_{tot} $ [10$^{28}$ erg]} \\
  \startdata
\multicolumn{2}{l}{Fit to RHESSI data only}  \\
1) \textit{thin\_kappa} & 	$31_{-9}^{+39}$&$4.1_{-0.1}^{+0.1}$&	$0.15_{-0.02}^{+0.01}$	& $14_{-2}^{+7}$ &$52_{-14}^{+31}$&$7.8_{-2.1}^{+4.7}$\\
2) $\xi_\kappa(T)$ &	$3_{-0.9}^{+11}$ &	$3.6_{-0.1}^{+0.1}$	&$0.17_{-0.07}^{+0.03}$		 &$4.5_{-0.7}^{+5}$ & $18_{-9}^{+28}$&$2.7_{-1.4}^{+4.2}$\\
\multicolumn{2}{l}{Fit to RHESSI and AIA data}  &  && &\\						
3a) $\xi_\kappa^{hot}(T)$  & $0.03_{-0.01}^{+0.01}$ & 	$3.9_{-0.1}^{+0.2}$&		$0.97_{-0.17}^{+0.04}$ & \multirow{2}{*}{$0.45_{-0.02}^{+0.02}$}& \multirow{2}{*}{$10_{-2}^{+1}$}	& \multirow{2}{*}{$1.5_{-0.3}^{+0.2}$}\\
3b) $\xi_\kappa^{cold}(T)$  &	$0.002_{-0.001}^{+0.001}$ 	&$4.5_{-1}^{+4}	 $&	$0.21_{-0.08}^{+0.01}$ &	 \\
   \enddata
  \end{deluxetable}
 \section{Discussion and conclusions}
We introduce a new representation of the kappa-distribution by means of a DEM and present a method that allows for fitting the new function to RHESSI data alone, as well as simultaneously with AIA data. In principle, any function that is solvable via Laplace transform can be fitted with this method \citep[see][]{Mo15}, but the kappa-distribution has a clear physical interpretation. The method is applied to a single-loop flare observed with RHESSI. While the RHESSI spectrum can be fitted well with a single kappa-distribution, inclusion of AIA data necessitates the addition of a second, low-temperature component. The overall spectrum of the analysed flare is consistent with a low-temperature ($\sim$ 2 MK) "core"-component and a high-temperature ($\sim$ 11 MK) kappa-distribution. This  is reminiscent of electron spectra observed in in-situ observations of the solar wind \citep[e.g.][]{1997AdSpR..20..645L,1997A&A...324..725M,2006LRSP....3....1M} which can often be fitted with a cooler core and a kappa-distribution. However, it has to be noted that the low-temperature component of the AIA DEM (dominant contribution to emission in 171 \AA\ and 211 \AA) is likely due to cool, foreground coronal plasma, as shown by \citet{2012ApJ...760..142B} and \citet{2014ApJ...780..107K} based on the temporal evolution of the DEM before and during flares. While this affects the interpretation of a core-plus-kappa-distribution of the flare accelerated electrons, it does not affect the total energy since the contribution of this cool component to the total electron number density and energy density is marginal. The calculations suggest that the previously used thin-target fit RHESSI spectra overestimates the number of electrons required to explain the X-ray and EUV emission by more than one order of magnitude, as suggested by Figure \ref{fig:DEM_nVF_combined}. \subsection{Total electron number density and flare energy}
The electron number density $n$ can be found from the observed emission measure and the volume via $n=\sqrt{EM/V}$. We approximate the volume as $V=A^{3/2}=1.5\times 10^{27}$ cm$^{3}$, using the area $A$ of the 50\% contour in the RHESSI 6--12 keV CLEAN image and accounting for the CLEAN beam-width. 
In the presented event, the total electron number density inferred from RHESSI data is $n=1.4\times 10^{11}$ $\mathrm{cm^{-3}}$ when the spectrum is fitted with \textit{thin\_kappa} and $n=4.5\times 10^{10}$ $\mathrm{cm^{-3}}$ when the X-ray spectrum is fitted with $\xi_\kappa(T)$. This constitutes a reduction by a factor of 3. When AIA data is added to further constrain the lowest energies, the resulting total number is $n=4.5\times 10^{9}$ $\mathrm{cm^{-3}}$ where both $\xi_{\kappa}(T)$ components were taken into account. This is a total factor of $\sim 30$ reduction compared with \textit{thin\_kappa} (see also Table \ref{tab:param}). Upper and lower limits were calculated using the upper and lower limits of the fit-parameters. 
In addition to the electron number density we can also calculate the total energy density from Equation (\ref{eq:meaneflux4}) as (see Appendix for full derivation):
\begin{equation}
U_\kappa=\frac{3}{2}k_BnT_\kappa,
\end{equation}
as well as the total energy by multiplying $U_\kappa$ with the volume (compare Table 1). Fitting RHESSI spectra with $\xi_\kappa(T)$ reduces the total energy  by a factor of $\sim$ 2.9 compared with \textit{thin\_kappa}. Simultaneously fitting RHESSI and AIA data $\xi_\kappa(T)$ gives a factor of $\sim$ 5 less energy than \textit{thin\_kappa}.  
This shows that, while fitting RHESSI spectra with $\xi_\kappa(T)$ already leads to a significant reduction of the total electron number, the lowest electron energies can only be truly recovered by simultaneously fitting RHESSI and AIA data, as can be seen in Figure \ref{fig:DEM_nVF_combined}.  

We conclude that: a) While the RHESSI data alone can be fitted with a single kappa-distribution in the presented flare, the inclusion of AIA data necessitates adding a second, low temperature component, most likely due to cool, non-flare related coronal plasma in the line of sight, which dominates the AIA emission in some wavelengths; b) fitting a thin-target kappa-distribution to RHESSI data leads to overestimation of the total number of energetic electrons needed to produce the observed X-ray emission by around a factor of 3 compared with fitting $\xi_\kappa(T)$; c) Simultaneous fitting of RHESSI spectra with AIA is necessary to properly constrain the low-energy part of the electron population, resulting in an overall reduction of factor $\sim$ 30 of the total electron number.  Therefore we encourage the use of the presented methodology to better estimate global flare energies.

\acknowledgments
MB was supported by the Swiss National Science Foundation (grant 200021-140308).
EPK gratefully acknowledges financial support by an STFC Grant. The work of GM was supported by programs P-9 and P-41 of the Presidium of the Russian Academy of Sciences, RFBR grants 13-02-00277A and 14-02-00924A.  EPK and GM acknowledge support by the Marie Curie International Research Staff Exchange Scheme "Radiosun" (PEOPLE-2011-IRSES-295272).

\bibliographystyle{apj}
\bibliography{mybib,all_issi_references}

\begin{thebibliography}{31}
\expandafter\ifx\csname natexlab\endcsname\relax\def\natexlab#1{#1}\fi

\bibitem[{{Battaglia} \& {Kontar}(2012)}]{2012ApJ...760..142B}
{Battaglia}, M., \& {Kontar}, E.~P. 2012, \apj, 760, 142

\bibitem[{{Battaglia} \& {Kontar}(2013)}]{2013ApJ...779..107B}
---. 2013, \apj, 779, 107

\bibitem[{{Bian} {et~al.}(2014){Bian}, {Emslie}, {Stackhouse}, \&
  {Kontar}}]{2014ApJ...796..142B}
{Bian}, N.~H., {Emslie}, A.~G., {Stackhouse}, D.~J., \& {Kontar}, E.~P. 2014,
  \apj, 796, 142

\bibitem[{{Brown} \& {Emslie}(1988)}]{1988ApJ...331..554B}
{Brown}, J.~C., \& {Emslie}, A.~G. 1988, \apj, 331, 554

\bibitem[{{Brown} {et~al.}(2010){Brown}, {Mallik}, \&
  {Badnell}}]{2010A&A...515C...1B}
{Brown}, J.~C., {Mallik}, P.~C.~V., \& {Badnell}, N.~R. 2010, \aap, 515, C1

\bibitem[{{Culhane}(1969)}]{1969MNRAS.144..375C}
{Culhane}, J.~L. 1969, \mnras, 144, 375

\bibitem[{{Culhane} \& {Acton}(1970)}]{1970MNRAS.151..141C}
{Culhane}, J.~L., \& {Acton}, L.~W. 1970, \mnras, 151, 141

\bibitem[{{Dere} {et~al.}(1997){Dere}, {Landi}, {Mason}, {Monsignori Fossi}, \&
  {Young}}]{1997A&AS..125..149D}
{Dere}, K.~P., {Landi}, E., {Mason}, H.~E., {Monsignori Fossi}, B.~C., \&
  {Young}, P.~R. 1997, \aaps, 125, 149

\bibitem[{{Dud{\'{\i}}k} {et~al.}(2012){Dud{\'{\i}}k}, {Ka{\v s}parov{\'a}},
  {Dzif{\v c}{\'a}kov{\'a}}, {Karlick{\'y}}, \&
  {Mackovjak}}]{2012A&A...539A.107D}
{Dud{\'{\i}}k}, J., {Ka{\v s}parov{\'a}}, J., {Dzif{\v c}{\'a}kov{\'a}}, E.,
  {Karlick{\'y}}, M., \& {Mackovjak}, {\v S}. 2012, \aap, 539, A107

\bibitem[{{Hannah} \& {Kontar}(2012)}]{2012A&A...539A.146H}
{Hannah}, I.~G., \& {Kontar}, E.~P. 2012, \aap, 539, A146

\bibitem[{{Holman} {et~al.}(2003){Holman}, {Sui}, {Schwartz}, \&
  {Emslie}}]{Ho03}
{Holman}, G.~D., {Sui}, L., {Schwartz}, R.~A., \& {Emslie}, A.~G. 2003, \apjl,
  595, L97

\bibitem[{{Holman} {et~al.}(2011){Holman}, {Aschwanden}, {Aurass}, {Battaglia},
  {Grigis}, {Kontar}, {Liu}, {Saint-Hilaire}, \&
  {Zharkova}}]{2011SSRv..159..107H}
{Holman}, G.~D., {Aschwanden}, M.~J., {Aurass}, H., {et~al.} 2011, \ssr, 159,
  107

\bibitem[{{Hurford} {et~al.}(2002){Hurford}, {Schmahl}, {Schwartz}, {Conway},
  {Aschwanden}, {Csillaghy}, {Dennis}, {Johns-Krull}, {Krucker}, {Lin},
  {McTiernan}, {Metcalf}, {Sato}, \& {Smith}}]{Hur02}
{Hurford}, G.~J., {Schmahl}, E.~J., {Schwartz}, R.~A., {et~al.} 2002, \solphys,
  210, 61

\bibitem[{{Inglis} \& {Christe}(2014)}]{2014ApJ...789..116I}
{Inglis}, A.~R., \& {Christe}, S. 2014, \apj, 789, 116

\bibitem[{{Ka{\v s}parov{\'a}} \& {Karlick{\'y}}(2009)}]{2009A&A...497L..13K}
{Ka{\v s}parov{\'a}}, J., \& {Karlick{\'y}}, M. 2009, \aap, 497, L13

\bibitem[{{Kontar} {et~al.}(2008){Kontar}, {Dickson}, \& {Ka{\v
  s}parov{\'a}}}]{2008SoPh..252..139K}
{Kontar}, E.~P., {Dickson}, E., \& {Ka{\v s}parov{\'a}}, J. 2008, \solphys,
  252, 139

\bibitem[{{Kontar} {et~al.}(2011){Kontar}, {Brown}, {Emslie}, {Hajdas},
  {Holman}, {Hurford}, {Ka{\v s}parov{\'a}}, {Mallik}, {Massone}, {McConnell},
  {Piana}, {Prato}, {Schmahl}, \& {Suarez-Garcia}}]{2011SSRv..159..301K}
{Kontar}, E.~P., {Brown}, J.~C., {Emslie}, A.~G., {et~al.} 2011, \ssr, 159, 301

\bibitem[{{Krucker} \& {Battaglia}(2014)}]{2014ApJ...780..107K}
{Krucker}, S., \& {Battaglia}, M. 2014, \apj, 780, 107

\bibitem[{{Landi} \& {Young}(2010)}]{2010ApJ...714..636L}
{Landi}, E., \& {Young}, P.~R. 2010, \apj, 714, 636

\bibitem[{{Landi} {et~al.}(2013){Landi}, {Young}, {Dere}, {Del Zanna}, \&
  {Mason}}]{2013ApJ...763...86L}
{Landi}, E., {Young}, P.~R., {Dere}, K.~P., {Del Zanna}, G., \& {Mason}, H.~E.
  2013, \apj, 763, 86

\bibitem[{{Lemen} {et~al.}(2012){Lemen}, {Title}, {Akin}, {Boerner}, {Chou},
  {Drake}, {Duncan}, {Edwards}, {Friedlaender}, {Heyman}, {Hurlburt}, {Katz},
  {Kushner}, {Levay}, {Lindgren}, {Mathur}, {McFeaters}, {Mitchell}, {Rehse},
  {Schrijver}, {Springer}, {Stern}, {Tarbell}, {Wuelser}, {Wolfson}, {Yanari},
  {Bookbinder}, {Cheimets}, {Caldwell}, {Deluca}, {Gates}, {Golub}, {Park},
  {Podgorski}, {Bush}, {Scherrer}, {Gummin}, {Smith}, {Auker}, {Jerram},
  {Pool}, {Soufli}, {Windt}, {Beardsley}, {Clapp}, {Lang}, \&
  {Waltham}}]{2012SoPh..275...17L}
{Lemen}, J.~R., {Title}, A.~M., {Akin}, D.~J., {et~al.} 2012, \solphys, 275, 17

\bibitem[{{Lin} {et~al.}(2002){Lin}, {Dennis}, {Hurford},
  {et~al.}}]{2002SoPh..210....3L}
{Lin}, R.~P., {Dennis}, B.~R., {Hurford}, G.~J., {et~al.} 2002, \solphys, 210,
  3

\bibitem[{{Lin} {et~al.}(1997){Lin}, {Larson}, {Ergun}, {McFadden}, {Carlson},
  {Phan}, {Ashford}, {Anderson}, {McCarthy}, {Skoug}, {Parks}, {R{\`e}me},
  {Bosqued}, {D'Uston}, {Sanderson}, \& {Wenzel}}]{1997AdSpR..20..645L}
{Lin}, R.~P., {Larson}, D.~E., {Ergun}, R.~E., {et~al.} 1997, Advances in Space
  Research, 20, 645

\bibitem[{{Maksimovic} {et~al.}(1997){Maksimovic}, {Pierrard}, \&
  {Lemaire}}]{1997A&A...324..725M}
{Maksimovic}, M., {Pierrard}, V., \& {Lemaire}, J.~F. 1997, \aap, 324, 725

\bibitem[{{Marsch}(2006)}]{2006LRSP....3....1M}
{Marsch}, E. 2006, Living Reviews in Solar Physics, 3, 1

\bibitem[{{Motorina} \& {Kontar}(2015)}]{Mo15}
{Motorina}, G.~G., \& {Kontar}, E.~P. 2015, Geomagnetism and Aeronomy, in press

\bibitem[{{Oka} {et~al.}(2013){Oka}, {Ishikawa}, {Saint-Hilaire}, {Krucker}, \&
  {Lin}}]{2013ApJ...764....6O}
{Oka}, M., {Ishikawa}, S., {Saint-Hilaire}, P., {Krucker}, S., \& {Lin}, R.~P.
  2013, \apj, 764, 6

\bibitem[{{Oka} {et~al.}(2015){Oka}, {Krucker}, {Hudson}, \&
  {Saint-Hilaire}}]{2015ApJ...799..129O}
{Oka}, M., {Krucker}, S., {Hudson}, H.~S., \& {Saint-Hilaire}, P. 2015, \apj,
  799, 129

\bibitem[{{Saint-Hilaire} \& {Benz}(2005)}]{2005A&A...435..743S}
{Saint-Hilaire}, P., \& {Benz}, A.~O. 2005, \aap, 435, 743

\bibitem[{{Smith} {et~al.}(2002){Smith}, {Lin}, {Turin}, {Curtis}, {Primbsch},
  {Campbell}, {Abiad}, {Schroeder}, {Cork}, {Hull}, {Landis}, {Madden},
  {Malone}, {Pehl}, {Raudorf}, {Sangsingkeow}, {Boyle}, {Banks}, {Shirey}, \&
  {Schwartz}}]{Smith02}
{Smith}, D.~M., {Lin}, R.~P., {Turin}, P., {et~al.} 2002, \solphys, 210, 33

\bibitem[{{Vasyliunas}(1968)}]{1968JGR....73.2839V}
{Vasyliunas}, V.~M. 1968, \jgr, 73, 2839

\end{thebibliography}

\appendix
\section{Derivation of total energy density}
The kappa-temperature $T_\kappa$ is commonly defined via the characteristic (i.e. most probable) speed, $\theta ^2 = \frac{2k_BT_{\kappa}}{m_e}\frac{\kappa -1.5}{\kappa}$. It has to be noted that, in our interpretation of the kappa-distribution as a summ of Maxwellians, $T_\kappa$ has the meaning of average temperature of the DEM. With this definition, the total energy density $U_\kappa$ and the average energy $\langle E \rangle$ can be expressed as $U_\kappa=\frac{3}{2}nk_BT_\kappa$ and $\langle E \rangle=\frac{3}{2}k_BT_\kappa$, respectively, as in the Maxwellian case. This can be shown straightforwardly by integration of the mean electron flux distribution:
\begin{equation}
\langle nVF(E)\rangle =C \frac{\frac{E}{k_BT_\kappa (\kappa-1.5)}}{(1+\frac{E}{k_BT_\kappa (\kappa-1.5)})^{(\kappa+1)}}
\end{equation}
with the constant
\begin{equation} \label{eq_a3}
C=\frac{n^2V 2^{3/2}\Gamma(\kappa+1)}{(\pi m_e)^{1/2}(k_BT_\kappa)^{1/2}(\kappa-1.5)^{3/2}\Gamma(\kappa-0.5)}
\end{equation}
gives
\begin{eqnarray}
U_\kappa&=&\int_0^\infty \langle nVF(E)\rangle \frac{E}{v}dE \quad \mathrm{[erg\, cm^{-3}]} \\
&=&\frac{C}{nV}\int_0^\infty \frac{\frac{E}{k_BT_\kappa (\kappa-1.5)}}{(1+\frac{E}{k_BT_\kappa (\kappa-1.5))})^{(\kappa+1)}}\frac{E}{\sqrt{2E/m_e}}dE.
\end{eqnarray}
Making the variable change $x=E/(k_B T_\kappa(\kappa-1.5))$ and solving the resulting integral using the Beta-function
\begin{equation}
\int_0^\infty\frac{x^{3/2}}{(1+x)^{\kappa+1}}dx=B\left(\frac{5}{2},\kappa-\frac{3}{2}\right)=\frac{\Gamma(5/2)\Gamma(\kappa-3/2)}{\Gamma(\kappa+1)}
\end{equation}
this results in:
\begin{equation}
U_\kappa=\frac{C}{nV}\sqrt{\frac{m_e}{2}}(k_BT_\kappa(\kappa-1.5))^{3/2}\frac{\Gamma(5/2)\Gamma(\kappa-3/2)}{\Gamma(\kappa+1)}
\end{equation}
Inserting Equation (\ref{eq_a3}) and using $\Gamma(\kappa-1/2)=\Gamma(\kappa-3/2+1)=(\kappa-3/2)\Gamma(\kappa-3/2)$ we find 
\begin{equation}
U_\kappa=\frac{2nk_BT_\kappa\Gamma (5/2)}{\pi^{1/2}}=\frac{3}{2}nk_BT_\kappa
\end{equation}
\end{document}